\documentstyle[12pt]{article}
\begin{document}
\begin{titlepage}

\title{Comments on the multi-dimensional Wheeler-DeWitt equation
         \thanks{Talk given at the International School-Seminar
                  ''Multidimensional Gravity and Cosmology'',
                  Yaroslavl, June 20-26, 1994}}

\author{Franz Embacher\\
        Institut f\"ur Theoretische Physik\\
        Universit\"at Wien\\
        Boltzmanngasse 5\\
        A-1090 Wien\\
        \\
        E-mail: fe@pap.univie.ac.at\\
        \\
        UWThPh-1994-36\\
        gr-qc/9409016}
\date{}

\maketitle

\begin{abstract}
It is argued heuristically -- using an
${\bf S}^3 \times {\bf S}^6$
minisuperspace model -- that there might be a fundamental quantum
gravity effect stabilizing internal spaces with non-vanishing
Ricci curvature.
\end{abstract}

\end{titlepage}

\section{Introduction}

In this contribution, I would like to present a particular speculation
in a quite heuristic manner. It concerns the stabilization of scale
factors belonging to internal spaces. The underlying physical
problem is provided by the typical classical behaviour of scale
factors in the very simplest higher-dimensional cosmological models
(see e.g. Refs.\cite{ChodosDetweiler,MatznerMezzacappa,somemore}):
some blow up, while the others usually collapse,
either approaching zero asymptotically
or running into a singularity after a finite amount
of proper time (''crack of doom'').
The papers \cite{ChodosDetweiler} and
\cite{MatznerMezzacappa}, as well as the references
contained therein provide examples for these two types.
Even those rare classical solutions which evolve towards finite
values of the internal
scale factors are unstable against small perturbations -- in contrast
to what we observe (clap your hands, and you will certainly not
cause an internal space to collapse).
\medskip

Stabilization of an internal space can be achieved in more
sophisticated models in which a particular interaction between
gravity and matter may capture some of the scale factors near a
(possibly local) minimum of an effective potential.
(Maybe the most popular approaches are those inspired by
higher-dimensional supergravity theories). For a selection of
some papers on this issue, see Refs.
\cite{Freund,stabmodels,Halliwellsix,CarowWatamuraetal,Wufour,Wudim}.
Some authors considered an effective
$R^2$-action for gravity \cite{ShafiWetterichReuter},
the use of finite temperature
quantum field theoretic methods (see e.g. Ref.\cite{Yoshimura}),
or cosmological models based on the theory of
superstrings \cite{strings}.
It is worth, however, thinking about the possible existence of
pure gravity mechanisms causing a comparable effect
in a logically and technically simpler way (by ''pure''
I mean that matter enters a model in a very simple or even a
more or less symbolic form, e.g. as a massive scalar field or
a cosmological constant). One such possibility might be
provided by the gravitational Casimir effect \cite{casimir},
but this is not very
much clear by now.
\medskip

Another area in which a ''pure gravity'' solution
to the stabilization problem
may be looked for is marked by the minisuperspace approaches to
quantum cosmology
\cite{HH,Hawkingscalar,qcminigeneral,Halliwell},
based on a (possibly path integral motivated) Wheeler-DeWitt equation
and boundary conditions for the wave function of the universe.
Concerning these attempts, one must say that only few solutions to few
models are known. There is especially one model -- maybe the most
natural to consider in this context -- that is remarkably resistent
against analytical methods: a ten-dimensional space-time with
${\bf S}^3 \times {\bf S}^6$
as spatial sections, and a positive cosmological constant $\Lambda$.
(I will definitely refuse to assume the existence of a
supergravity-inspired rank six antisymmetric tensor field which
would generate stability of the ${\bf S}^6$ by virtue
of a Freund-Rubin mechanism \cite{FreundRubin}; see also
Ref. \cite{CarowWatamuraetal}).
This model I will talk about, sometimes using
${\bf S}^3 \times {\bf T}^6$
for comparison.
\medskip

By the shorthand notations used above I mean the class of
metrics
\begin{equation}
ds^2 = - {\cal N}(t)^2 dt^2 + a_1(t)^2 d\sigma_3^2 + a_2(t)^2 d\sigma_6^2
\label{metric}
\end{equation}
where $d\sigma_3^2$ is the metric on the round unit three-sphere, and
$d\sigma_6^2$ is the metric on the round unit six-sphere (or on the
flat unit six-torus, in which case the internal dimensions do not
contribute a curvature term in the action). This model (and,
more general,
${\bf S}^m \times {\bf S}^n$
with $m,n \geq 2$) showed up in some of the lists given by other
contributors at this conference (e.g. Sascha Zhuk),
and it was implicitly
declared "non-fully-integrable" in the sense of not being reducible
to an (integrable) Toda system. We shall see that this
characterization is possibly connected with physically desirable
properties. The mathematical problem is that one has to treat
the two scale factors $a_1$ and $a_2$ on essentially the same
footing -- there seem to be very few possibilities to get rid of one
degree of freedom (e.g. by perturbative approximations). This is
in contrast, for example, to the Friedmann-Robertson-Walker (FRW)
models with a single scalar
field $\phi$ \cite{Hawkingscalar},
where, in a first approximation, $\phi$ appears only
through the effective cosmological constant $V(\phi)$, and the
two dimensions of minisuperspace are reduced to one. As a consequence,
the application to (\ref{metric}) of methods developed for simpler
models
\cite{Halliwellsix,HH,Hawkingscalar,qcminigeneral,Halliwell}
is likely to be problematic, and
we are far from knowing the set of all physically reasonable
quantum states. Some material on this model may be found in
Refs. \cite{HuWu,Wufour,Wudim,CarowWatamuraetal}).
\medskip

In order to develop my speculation, let me outline briefly what one
usually does in minisuperspace quantum cosmologies, using (\ref{metric})
as a starting point. (For an excellent introduction to this subject
see Ref.\cite{Halliwell}.)
The Einstein-Hilbert action for
ten-dimensional gravity, including a (positive) gravitational constant and
the usual boundary term that subtracts a divergence, is given by
\begin{equation}
S = - C \int_{{\cal M}_{10}} d^{\,10}x \sqrt{-g}
    (\,{}^{10}R + 2\, \Lambda)
    - 2\, C \int_{\partial {\cal M}_{10}} d^{\,9} x \sqrt{h} K,
\label{EinsteinHilbert}
\end{equation}
where
\begin{equation}
C = \frac{m_P^2}{16 \pi} (\rm{volume\, of\, internal\, space\, today})^{-1},
\end{equation}
$h_{ij}$ the metric induced by $g_{\mu\nu}$ on the boundary
$\partial {\cal M}_{10}$
and $K$ the trace of its extrinsic curvature.
The above choice of $C$ ensures the correct gravitational
constant today.
Inserting (\ref{metric}), one obtains an action
$S = \int dt {\cal L}(a_1,a_2,\dot a_1,\dot a_2,{\cal N})$
that we will not display (see e.g. Ref.\cite{Yoshimura}),
but just mention that the total volume of the $t = const$ space
sections is given by
\begin{equation}
V = \int_{{\cal M}_9}d^{\,9} x\sqrt{h} = w a_1^3 a_2^6,
\end{equation}
where $w=32 \pi^5/15$ in the ${\bf S}^6$ case (then we define
$k=1$), and $w=2^7 \pi^8$ in the ${\bf T}^6$ case ($k=0$).
\medskip

Next we change variables according to
\begin{eqnarray}
N & = & \frac{\Lambda}{3 \sqrt{10}}\, a_2\, {\cal N}\nonumber\\
u & = & \frac{\Lambda^2}{180}\, a_1^2\, a_2^2\label{transf}\\
v & = & \frac{\Lambda^3}{5400 \sqrt{5}}\, a_1 \,a_2^5\nonumber\\
\nonumber
\end{eqnarray}
and make the definition (using $\Lambda \equiv \ell_{\Lambda}^{-2}$
and $m_P \equiv \ell_{P}^{-1}$)
\begin{equation}
\Lambda_{\rm {eff}} = \frac{375^{1/4} \Lambda}{2^{1/8}\,6\, (w C)^{1/4}}
                   \equiv \Bigg(
                   \frac{1}{\pi}
                   \bigg(\frac{5}{3\sqrt{2}}\bigg)^3
                   \bigg (\frac{\ell_P}{\ell_{\Lambda}}\bigg)^2
                   \bigg(\frac{a_2(\rm{today})}{\ell_{\Lambda}}\bigg)^6
                                        \Bigg)^{1/4},
\label{Lambdaeff}
\end{equation}
the last identity being valid only in the $k=1$ case
when $a_2$ compactifies.
(Let us remark here that one could consider any
${\bf S}^n$
($n \geq 2$) instead of ${\bf S}^6$, but then the prefactors
and exponents in these definitions would be even more messy).
In these new variables, the action (\ref{EinsteinHilbert})
becomes
\begin{equation}
S = \frac{1}{\Lambda_{\rm{eff}}^4}
   \int dt \left( -\frac{\dot u \dot v}{N}
                 - N W(u,v) \right),
\label{S}
\end{equation}
where the potential is given by
\begin{equation}
W(u,v) = - v - k\, u^{3/2} + 2 u^{5/4} v^{1/2}.
\label{W}
\end{equation}
The corresponding Euclidean action (obtained by sending
$t \rightarrow i t$,
thus changing (\ref{metric})
into the Riemannian metric
$ds^2_{E}={\cal N}(t)^2 dt^2+a_1(t)^2 d\sigma_3^2+a_2(t)^2 d\sigma_6^2$)
reads
\begin{equation}
I = \frac{1}{\Lambda_{\rm{eff}}^4}
   \int dt \left( -\frac{\dot u \dot v}{N}
                 + N W(u,v) \right).
\label{I}
\end{equation}
The Wheeler-DeWitt metric reduces to $ds^2_{WDW} = - du dv$, which shows
that $u$ and $v$ are ''lightlike'' coordinates in
minisuperspace (both ranging from $0$ to $\infty$).
We define the ''timelike'' direction in minisuperspace by
$du dv > 0$.
The Hamiltonian corresponding to (\ref{S}) is the
well-known energy constraint
\begin{equation}
{\cal H} = N \left( - p_u p_v + \Lambda_{\rm eff}^{-8}\,\, W(u,v) \right)
\label{H}
\end{equation}
that has to vanish (on the constrained phase space or on physical
quantum states, respectively). The classical (Lorentzian) solutions
of this system must obey (in the gauge $N=1$)
\begin{equation}
\dot{u}\dot{v}=\pm \, W, \qquad
\ddot{u} = \pm \, \partial_v W, \qquad
\ddot{v} = \pm \, \partial_u W,
\label{eqmotion}
\end{equation}
with the + signs (the first of these just meaning ${\cal H}=0$);
the corresponding Euclidean solutions -- following from variation
of $I$ -- obey (\ref{eqmotion}) with the -- signs. Lorentzian and
Euclidean solutions (trajectories in minisuperspace) describe
ten-dimensional geometries when re-transformed and inserted into
(\ref{metric}) and its Euclidean counterpart, respectively.
The variables chosen are such that the form of the
classical equations of motion is formally independent of $\Lambda$
and $C$.
\medskip

The transition to quantum mechanics is achieved by replacing
$p_u \rightarrow - i \partial_u, \,
p_v \rightarrow - i \partial_v$
in $\cal H$, thus leading to the Wheeler-DeWitt equation
\begin{equation}
\partial_{uv} \psi(u,v) = - \Lambda_{\rm eff}^{-8}\,\, W(u,v)\,\, \psi(u,v).
\label{WDW}
\end{equation}
The operator ordering has been chosen in the only way that is
consistent with general covariance under transformations of all
three variables, including the lapse function \cite{Halliwell}.
(Let us note in
brackets that by choosing other variables
$\bar u=\bar u(u),\,\bar v=\bar v(v)$, one can achieve various
forms of the potential
$\bar W=W \frac{du}{d\bar u} \frac{dv}{d\bar v}$
by transforming the lapse as
$\bar N=N \frac{d\bar u}{du} \frac{d\bar v}{dv}$. However,
the choice (\ref{W}) seems to me to be the best one.
\medskip

Having set up convenient variables, let us now look at some
basic features of the model. (Clearly, in view of (\ref{eqmotion},
\ref{WDW}), all such features are contained in the potential $W$,
because the rest is fairly trivial). First we note that there
is a curve $W=0$ (which may be found by solving (\ref{W})
with respect to $v$), lying entirely in the interior of
minisuperspace (see Fig.1).
In the region $W<0$ (which extends to the axes
$u=0$ and $v=0$), the Euclidean solutions are
necessarily timelike curves (with respect to the Wheeler-DeWitt
metric), the Lorentzian ones are spacelike
curves (cf. the first equation of (\ref{eqmotion})).
The opposite is of course true if $W>0$.
An escape of both scale factors $(a_1,a_2)$ to large values is
only possible inside the $W>0$ region which one is tempted to call
the ''classical'' one. The region $W<0$ could then be called the
''quantum'' or ''Euclidean'' regime. Note that the two
asymptotic branches of the zero-potential curve
($v \leadsto 4 u^{5/2}$ for the ''upper'' and
$v \leadsto \frac{1}{4} u^{1/2}$ for the ''lower'' branch
in Fig.1)
approach
constant $a_1 \rightarrow \ell_{\Lambda} \sqrt{3}$ and
$a_2 \rightarrow \ell_{\Lambda} \sqrt{15}$, respectively (which is, if
$\ell_{\Lambda} \approx \ell_P$,
what one expects from a classical/quantum transition regime).
\medskip

Moreover, the curve $W=0$ provides all points in which a
classical Euclidean solution may be matched smoothly to a
Lorentzian one.  It is not quite clear which role
Euclidean trajectories should play in a sensible quantum cosmology,
but in some of the most prominent approaches
they provide
a key of finding the state of the universe. Euclidean
trajectories describing regular ten-geometries emerge from the
origin ($u(0)=v(0)=0$) and behave there like
$v \sim c_1 u^{5/2}\,({\rm then\,\,} a_1(0)={\rm finite})$ or
$v \sim c_2 u^{1/2}\,({\rm then\,\,} a_2(0)={\rm finite})$.
\medskip

One usually singles out two particluar Euclidean solutions as
''instantons'': one being along the curve $v=(9/16) u^{5/2}$
(in $W<0$) and the other along $v=(4/9)u^{1/2}$ (in $W<0$, too).
They give rise to the Riemannian ten-geometries
${\bf S}^3 \times {\bf S}^7$ and
${\bf S}^4 \times {\bf S}^6$,
respectively and are the only Euclidean
solutions starting at the origin and having a turning point
($\dot u=\dot v=0$, i.e. zero extrinsic curvature) at $W=0$.
They are usually considered the preferred
candidates describing the coming-into-(classical)existence
of the universe by quantum tunneling
\cite{CarowWatamuraetal,Wufour,Wudim,HuWu}.
In Fig.1, these
instantons as well as several other typical Euclidean
trajectories are displayed.
\medskip

When expressed in terms of their proper-time (${\cal N}=1$),
the Lorentzian solutions either expand in both scale factors
$a_1$ and $a_2$ exponentially
(thus describing inflation
without compactification), or they expand in one scale factor
and contract in the other, thereby entering the ''quantum''
region $W<0$ and finally collapsing towards one of the axes
in a Kasner-type singularity ($a_1 \rightarrow \infty$ while
$a_2 \rightarrow 0$, or vice versa).
The only solutions that compactify
are the Lorentzian analogues
of the two instantons (lying on the same curves as these,
but now in the region $W>0$),
and even these are unstable against
small perturbations.
Each of these two very special solutions are characterized by
one scale factor being actually constant ($a_1=\ell_\Lambda \sqrt{8}$
and $a_2=\ell_\Lambda \sqrt{20}$, respectively). In the
approaches that connect the instantons to the most
probable classical evolution of the universe
\cite{CarowWatamuraetal,Wufour,Wudim,HuWu},
this last number is
interpreted as the actual Kaluza-Klein scale factor.
In Fig.2, some Lorentzian trajectories are
shown.
\medskip

Comparing this structure to the $k=0$ case, we find that there
the zero-potential curve $W=0$ is just given by $v=4 u^{5/2}$,
and the lower asymptotic branch has disappeared.
\medskip

Can the transition from the classical trajectories satisfying
(\ref{eqmotion}) to the Wheeler-DeWitt equation (\ref{WDW})
improve likelihood and stability of compactification?
Let me as a starting point adopt the prescription of Hartle
and Hawking \cite{HH}
for finding the ''no-boundary''-solution
of (\ref{WDW}). In this approach, the wave function is
considered as a path integral
$\psi = \int {\cal D}g\, \exp(-I[\,g\,])$ over compact Euclidean
ten-geometries having the argument of $\psi$
(a nine-geometry or, here, just a point $(u,v)$) as their
boundary. The WKB-approximation scheme \cite{Halliwell}
then tells us to
compute $\psi \approx A \exp(-I)$ for points $(u,v)$
near (or on) the zero-potential curve, where the
Euclidean action $I(u,v)$ is taken along the Euclidean trajectories.
$A$ is  a prefactor that may be estimated by WKB-methods,
once $I$ is known.
We would then have to take this $\psi(W=0)$ as a boundary
condition for the Wheeler-DeWitt equation (\ref{WDW}) and
evolve it into the region $W>0$. There, $\psi$ is expected
to develop a form
$\psi_{WKB} \approx B \cos(S) \equiv \frac{1}{2}B(\exp(i S)+\exp(-i S))$.
The phase $S(u,v)$ is the action of a family of classical solutions
that one may find using the Hamilton-Jacobi formalism \cite{Halliwell}.
The probability measure is provided by the prefactor $B$,
and $\psi$ can be viewed as describing a superposition of a family
of classical universes (a particular one of course being
ours). Some of these universes inflate in both scale factors, and
some will eventually re-enter the quantum region. Only two
classical paths (namely the ones emerging from the two instantons
by smooth continuation) will undergo compactification of
one scale factor
-- and although $\psi$ can be expected
to be highly peaked around these (as dominant contributions
to the path integral), the situation is unsatisfactory as far
as stability is concerned. (Remember that I refuse the introduction
of a stabilizing supergravity-inspired matter coupling).
\medskip

However, one encounters some complications in this model:
To begin with, let us note that the
Euclidean action $I(u,v)$ is not unique, since there are several
Euclidean trajectories connecting a given $(u,v)$ with the
origin. (Choosing the largest or the smallest of these would
presumably not result into an approximate solution of the
Wheeler-DeWitt equation at all, due to the effects at those
points where two trajectories give equal $I$).
\medskip

Moreover, the interpretation of the region $W<0$ according to
the standard methods is questionable: Since in a
large portion of this region the curves $W=const$ are timelike
with respect to the Wheeler-DeWitt metric, $\psi$ would
have to be expected oscillatory there rather than exponential
\cite{Halliwellsix,Halliwell}.
This is however contrary to one's intuition about what should
happen when internal spaces approach Planck scale size.

\section{Speculation}

The question I would like to focus on is the following:
What happens to the universe if it is described in the classical
($W>0$) region by a trajectory that approaches the
zero-potential curve, enters the region of negative $W$ and -- when
evolved further classically -- recollapses? If $\psi$ in the
region $W<0$ (for $u$ and $v$ large) is indeed a semi-classical
state, one should expect the universe to run into a singularity.
There are some remarks about similar situations
in the literature,
stating either that the universe actually {\it will} recollapse
classically \cite{Hawkingscalar}
or that there might possibly occur some kind of collapse
by tunneling \cite{Halliwellsix}.
In any case, such trajectories are usually not considered
as important in the description of our universe.
\medskip

Now, let us proceed (very) heuristically and retain -- against
possible objections -- the notion that a physically sensible state
has to qualify the region $W<0$ as a ''quantum'' or a
''classically forbidden'' domain (at least if
$\ell_\Lambda \approx \ell_P$), irrespective of the fact
that classical solutions do exist there formally.
Then the question is: What happens, if the universe re-enters
into a classically forbidden region of superspace?
The intuitive answer is: tunneling. Since I have no better recipe
at hand (maybe a path integral formalism would provide one),
and since the Euclidean trajectories are in general considered
as a viable tool describing tunneling,
I try to {\it match a Euclidean trajectory smoothly to the re-entering
Lorentzian one}. This Euclidean trajectory will in general return
to the curve $W=0$ and generate another semiclassical
universe (which might recollapse again and so forth).
Hence, we arrive at a procedure which tells us to evolve
trajectories according to the classical equations of
motion (\ref{eqmotion}), but with the respective signs in each
region. These (mixed) trajectories correspond to metrics undergoing
an infinite succession of signature changes.
\medskip

There are certainly many reasonable objections against such a
procedure playing any role in the interpretation
(or construction) of a state $\psi$.
One such objection is provided by
the fact that just matching trajectories smoothly as described
above does not automatically lead to smooth ten-geometries
displaying signature change (nor to a stationary point of
the action $S$ or $I$).
Optimal matching would rather demand
vanishing extrinsic curvature \cite{signaturechange1} at $W=0$
(and this is in turn only possible for the two instantons, due to
their turning point structure).
But even this point is not so clear, because the recent discussion on
signature change shows that weaker junction conditions may allow
for reasonable classical evolution \cite{signaturechange2} too.
I will return to this in the very last remark of my talk.
However, even if the mixed trajectories do not represent
admissable geometries,
it may well be that they are linked with
dominant contributions to the path integral with
respect to some measure. (Recall that the structure of Euclidean
trajectories in the minisuperspace of the model we consider
is rather involved, and
to find the dominant ones for a given point $(u,v)$
is not a simple matter,
even in the  ''standard'' Hartle-Hawking approach
\cite{HH}).
Moreover, it is of course necessary to ask for the significance of
the ''time'' variable $t$ along the Euclidean pieces, and in which
way such a construction is linked to physical predictions.
\medskip

However, there might be a more accurate version of this procedure,
where the matching of trajectories is not understood
individually (one would expect a whole family of such mixed
trajectories to build up a quantum state anyway). The weakest
question one may make out of this is: Is there a solution of
the Wheeler-DeWitt equation (\ref{WDW}) which ''drags'' the
universe approximately along such paths? This could happen
either in terms of some oscillating WKB-type wave function
(which is less likely) or
in terms of a non-oscillating $\psi$ displaying huge
amplitudes in the according region of minisuperspace.
(Recall that -- except in the semiclassical WKB-approximation --
there is no viable and commonly accepted interpretation of
solutions to (\ref{WDW}), especially when such high peaks
occur in the absence of oscillations.
This touches upon the most fundamental
conceptual problems of quantum gravity \cite{AshtekarStachel},
and therefore I cannot even give a precise statement which mathematical
properties such a wave function should have.)
\medskip

It is time now to look at the particular way, the
Euclidean/Lorentzian mixed trajectories behave in the model
we are considering. Using simple numerical techniques, it
turns out that there is a preferred region on the curve $W=0$
from which such trajectories emerge (this region seems to consist
of two pieces which are located near the turning points of the
two instantons, and consist of points such that
two different Euclidean trajectories starting from the
origin and meeting there have approximately the same
action $I$).
\medskip

Fig.3 and Fig.4  show two examples for mixed trajectories.
They undergo ''oscillations'', thereby wandering along
one of the two asymptotic branches of $W=0$. One of the two
scale factors thereby approaches a finite limiting value in
each case, the other one blows up. In Fig.5, the data
from Fig.4 are re-expressed in terms of $(a_1(t),a_2(t))$,
where the time coordinate $t$ refers to the gauge $N=1$.
Fig.6 provides a magnification of the same plot, showing that
the ''oscillation'' of $a_2$ is actually a dampted one.
Hence, mixed trajectories correspond to metrics that
''oscillate'' in signature and display -- as far as the values
of the scale factors are concerned -- perfect Kaluza-Klein
behaviour.

\section{Outlook}

So, can the universe ''reappear'' due to quantum tunneling?
It is a subject for further research to look for solutions
of (\ref{WDW}) describing such a behaviour. Let me however
conclude with some remarks. First, such a compactification
and stabilization mechanism would be quite general and independent
of most features of matter coupling. It just requires the
internal space to have non-vanishing Ricci curvature.
(Note that in the
${\bf S}^3 \times {\bf T}^6$
model compactification of the ${\bf T}^6$
cannot happen in this way
because the lower asymptotic branch of $W=0$
is missing). Moreover, the compactified scale factor
approaches $\approx \ell_\Lambda$, hence the whole mechanism
works only if $\Lambda$ does not decay. It must be a fundamental
part of the action (and not just mimicking a $V(\phi)$ which
would disappear after an inflationary phase \cite{Hawkingscalar}).
A possible success of computations along these lines
could certainly be connected with a new aspect in the
interpretational problems of the state $\psi$,
because in some sense the universe would remain permanently
a quantum one.
One would have to work out how the arrow of time comes
about, and how the universe as a whole is experienced.
Naively, one would expect that a state in which $a_2$
compactifies (cf. Fig.5) describes an effectively
three-dimensional world expanding as $a_1(t)$. But why do
we observe a classical space-time with Lorentzian signature,
and how does $a_1$ evolve with respect to an (effectively)
Lorentzian time?
\medskip

Let me add a comment that emerged in discussions at this
meeting: When trying to compute $I(u,v)$ along
Euclidean trajectories, one would -- as already mentioned --
encounter different, competing contributions
$I_1(u,v)$ and $I_2(u,v)$, say. This degree of non-uniqueness
can -- by virtue of rather involved branching phenomena --
result into the mathematical structure necessary to develop
non-expected compactification solutions and at the same time
into the feature of being ''non-integrable'' in some sense.
Hence, the difficulties in finding analytic solutions
may possibly be compensated by new physical effects.
In this concern, there is also still something to learn
about the $k=0$ model (where the
${\bf S}^3$
would compactify along the lines described above).
\medskip

My last remark concerns the form of the Wheeler-DeWitt equation
(\ref{WDW}). One might consider the Euclidean solutions as
physical (i.e. classical) ones (and not just tools to integrate
(\ref{WDW}) or to evaluate a path integral). In other words,
a physical signature change
would actually be a thing to happen.
In this case, it is worth noting that the mixed trajectories
extremize the action $S_{\rm mod}$, obtained from
(\ref{S}) after replacing $W$ by its
absolute value $|W|$. To what extent they represent
physical solutions is not totally clear, because the question
which junction conditions should apply
at the surface of signature change
is still a bit open \cite{signaturechange1,signaturechange2}.
In any case, the ''classical signature change'' approach
favours weaker conditions (continuity of the extrinsic curvature
instead of vanishing), and the mixed
trajectories I described are likely to be accepted as
classically sensible solutions. Clearly, such an approach
alters the standard foundations of quantum gravity.
Following the spirit of my
speculations, one would modify the Wheeler-DeWitt equation
accordingly, and thus arrive at
\begin{equation}
\partial_{uv} \psi(u,v) = - \Lambda_{\rm eff}^{-8}\,\,
                  | W(u,v)|\,\, \psi(u,v)
\label{WDWmod}
\end{equation}
instead of (\ref{WDW}).
Expressed briefly, treating signature change as a classical
effect, the substitution
$p_u \rightarrow - i \partial_u, \,
p_v \rightarrow - i \partial_v$
in the Euclidean Hamiltonian
${\cal H}_E = N ( - p_u p_v - \Lambda_{\rm eff}^{-8}\,\, W )$
leads to an equation like (\ref{WDW}) but with
$W \rightarrow -W$ replaced. In contrast, the usual quantum
cosmology approach amounts to perform the
Euclideanized substitution
$p_u \rightarrow \partial_u, \,
p_v \rightarrow \partial_v$
in ${\cal H}_E$, which leads back to (\ref{WDW}). The statement
that the use of either Lorentzian or Euclidean geometries
makes no difference in deriving the Wheeler-DeWitt equation
\cite{Hawkingscalar} is only true if the latter are of no
classical significance. Postulating Euclidean signature in
the ''quantum domain'' of minisuperspace (which may well be
a ''classical'' domain in this context, especially if
$\ell_\Lambda \gg \ell_P$), leaves the zero-potential
curve as the set of points where signature change {\it may}
happen. Assuming further that it {\it does} happen there,
one arrives at (\ref{WDWmod}).
\medskip

A modified Wheeler-DeWitt equation as in (\ref{WDWmod})
is exactly what has been proposed recently by Martin \cite{Martin}.
In his paper, Martin also shows
(in a FRW plus scalar field minisuperspace model)
that the modified Wheeler-DeWitt
equation allows less physically sensible boundary conditions
than the standard one. If such an approach turns out to be
fruitful for quantum cosmology, the speculations made here
would turn out to be much less speculative than in the usual
context, but would simply be a heuristic description of a
particular candidate semiclassical state
$\psi \sim \Sigma \cos(S_{\rm mod})$ or
$\Sigma \exp(i S_{\rm mod})$.
Maybe the following (very last) speculation is true: states
built around mixed trajectories appear in the standard
cosmological approach as highly peaked in the $W \approx 0$
region of superspace, whereas in the classical signature
change picture they emerge as oscillating wave functions.
\\
\\
\\
{\Large {\bf Acknowledgments}}
\medskip

This work was supported by the Austrian Academy of Sciences
in the framework of the ''Austrian Programme for Advanced Research
and Technology''. Also, I would like to express my thanks to the
organizers of this meeting.

\bigskip
\bigskip
\bigskip

\noindent
{\Large {\bf Figure captions:}}
\\
\\
\\
{\large {\bf Fig.1}}\\
This plot displays, using a $(u,v)$-coordinate system of minisuperspace,
the zero-potential curve (dashed), with its two asymptotic branches
that are called ''upper'' and ''lower'' in the text, as well as some
classical Euclidean trajectories. The instantons are represented by
those two curves which appear to end at $W=0$ (they actually have turning
points there). The arrows indicate the direction off the
regular zero-geometry.
Recall that the ''timelike'' direction in minisuperspace is given
by $du dv>0$. This may be considered defining a ''light-cone''
that is rotated by $45^\circ$ against the usual one
in the special relativistic $(t,x)$-diagrams.
\\
\\
{\large {\bf Fig.2}}\\
Some Lorentzian (''physical'') trajectories are drawn. The
only solutions describing compactification (i.e. the counterparts
of the two instantons) are those curves which end at their
turning points on
the (dashed) zero-potential line.
All other trajectories displayed were (without loss of
much generality) chosen without
turning point and are extended maximally in both
directions of their time parameter.
Concerning the fact that
some of them hit the axes, recall from (\ref{transf})
that, for example, $v \rightarrow 0$ at finite $u$
means $a_1 \rightarrow \infty$
while $a_2 \rightarrow 0$. This behaviour is referred to as
''Kasner-type'' singularity in the text. A nice example
is provided by the trajectory starting from $(u \approx 6.79,v=0)$,
evolving into the region $W>0$, re-entering the
$W<0$ domain at $(u \approx 5,v \approx 0.62)$ and
running back into the $u$-axis at $u \approx 20.85$ (outside
the plot). A similar fate occurs to the trajectory starting
at $(u=0,v \approx 0.92)$ which re-enters the $W<0$ domain at
$(u \approx 11.1,v \approx 0.87)$
and finally collapses towards the $u$-axis at $u \approx 46.4$.
\\
\\
{\large {\bf Fig.3}}\\
A particular trajectory changing its character between
Euclidean and Lorentzian (denoted ''mixed trajectory'' in
the text) is displayed. It ''oscillates''
along the upper branch of the $W=0$ curve (dashed) and
hence approaches asmmptotically a constant value
$a_1 \rightarrow \ell_\Lambda \sqrt{3}$. The asymptotic
boundary condition near the origin (from where the
numerical evolution starts) has been chosen as
$u^{-5/2} v \rightarrow 0.25$.
\\
\\
{\large {\bf Fig.4}}\\
This plot shows another mixed trajectory, now ''oscillating''
around the lower branch of the zero-potential curve.
The scale factor belonging to ${\bf S}^6$ approaches
$a_2 \rightarrow \ell_\Lambda \sqrt{15}$. The asymptotic
behavour near the origin is
$u^{-1/2} v \rightarrow 0.5$.
\\
\\
{\large {\bf Fig.5}}\\
The graphs of the functions
$a_1(t)$ and $a_2(t)$ for the trajectory displayed in
Fig.4 are shown. The gauge condition defining $t$
is $N=1$. In order to get definite units,
we have chosen $\Lambda=1$, hence $\ell_\Lambda=1$.
\\
\\
{\large {\bf Fig.6}}\\
This provides a magnification of Fig.5 in order to
show how $a_2(t)$ performs ''damped'' oscillations
arounds its limiting value.

\end{document}